\documentclass[10pt,preprint]{aastex}
\usepackage{graphicx}

\begin{document}

\title{The 2dF Redshift Survey II: UGC 8584 - Redshift Periodicity and Rings}

\author{H. Arp}


\affil{Max-Planck-Institut f\"ur Astrophysik, Karl
 Schwarzschild-Str.1, Postfach 1317, D-85741 Garching, Germany}
 \email{arp@mpa-garching.mpg.de}

\author{C. Fulton}
\affil{Centre for Astronomy, James Cook University, Townsville
Queensland, 4811, Australia}
\email{mainseq@thevine.net}

\begin{abstract}

UGC 8584 was selected by a computer program as having a number of
quasars around it that obeyed the Karlsson periodicity in its
reference frame. On closer examination 9 of the nearest 10 quasars
turned out to be extremely close to the predicted values. Also it
turned out that UGC 8584 was a disturbed triple galaxy and a strong
triple radio source as well as being a strong millimeter and infrared
source. Evidence for present ejection velocities of $z_v \sim .01$
for the associated quasars is present and some pairing of ejections is
noted.

A new and important result emerges from this sample of galaxy/quasar
families, namely that rings and shells of galaxies and quasars tend to
surround galaxies which have active nuclei. Test cases suggest
obscuration of the background around these galaxies out to about $20'$ or
beyond. Because incidents of strong reddening are not observed, obscuring
particles are suggested to be large compared to optical wavelengths. In
principle, material ejected with the quasars could be of sizes of
gravel, boulders or larger.

\end{abstract}


\section{Introduction}

In a computer analysis of the 2dF redshift survey, groups of quasars
that obeyed the Karlsson values with respect to neighboring galaxies
were catalogued in Fulton \& Arp (2006) (Paper I). UGC 8584 turned out to have 9
of its nearest quasars fall especially close to the standard Karlsson
values. Figure 1 and Table 1 show that 9 of the nearest 10 have an
average periodicity residual of only ${|z_v|} \ge .010$. Since there is a
range of about $.105$ between adjacent periodicity peaks in $z_v$, each
of these $z_v$'s has a chance of about $z_v/.0525$ of being accidental. The
highest $z_v$ of the first 9 is $.0165$. Therefore the chance of 9 of the
first 10 falling by accident this closely to the peaks is

           $$P = 10!/9! \times 0.29^9 \times .72 = 1.0 \times 10^{-4}$$

The effect shown in Figure 1 is so strong that the cumulative average $z_v$
never reaches its expected value of $.0525$ for random $z$'s even at the
edge of the field ($r = 30'$). For this particular galaxy-quasar family
then there appears to be strong evidence that the quasar redshifts
obey the Karlsson series first annunciated in 1971 (Karlsson 1971).

\begin{table}[ht]
\caption{Quasars Near UGC 8584} \label{QuasarsNearUGC8584}
\vspace{0.3cm}
\begin{tabular}{lcccc}
{\bfseries Object\/} & {\bfseries $z$\/} & {\bfseries $r'$\/} & {\bfseries $z_0$\/} & {\bfseries $z_v$\/} \\
\hline
UGC 8584 & 0.060 &  0.0 & ...  & ...         \\
2QZ      & 1.583 &  3.4 & 1.44 & +.011$^{a}$ \\
2QZ      &  .714 &  7.5 &  .62 & +.011$^{a}$ \\
2QZ      & 2.168 &  8.4 & 1.99 & +.010$^{a}$ \\
SDSS     & 1.807 &  8.9 & 1.65 & +.099$^{b}$ \\
2QZ      & 1.542 & 10.4 & 1.40 & -.005$^{a}$ \\
LBQS     &  .672 & 12.0 &  .58 & -.014$^{a}$ \\
2QZ      &  .668 & 15.0 &  .57 & -.017$^{a}$ \\
LBQS     & 2.800 & 15.7 & 2.59 & -.015$^{a}$ \\
2QZ      & 2.189 & 16.5 & 2.01 & +.016$^{a}$ \\
SDSS     &  .702 & 16.7 &  .61 & +.003$^{a}$$^{b}$ \\
$$       &       &      &      &        \\
2QZ      & 2.547 & 17.4 & 2.35 & -.09  \\
2QZ      & 2.168 & 19.8 & 1.99 & +.01  \\
2QZ      & 1.918 & 20.6 & 1.75 & -.07  \\
SDSS     & 2.027 & 21.5 & 1.86 & -.03  \\
2QZ      &  .687 & 23.6 &  .59 & -.01  \\
2QZ      &  .829 & 23.9 &  .73 & +.08  \\
2QZ      & 1.504 & 27.1 & 1.36 & -.02  \\
SDSS     & 1.858 & 28.6 & 1.70 & -.09   \\
2QZ      & 1.646 & 28.6 & 1.50 & +.04  \\
2QZ      & 2.438 & 28.8 & 2.24 & +.10  \\
2QZ      &  .939 & 28.8 &  .83 & -.07  \\
2QZ      & 2.277 & 29.4 & 2.09 & +.04  \\
2QZ      & 2.093 & 29.4 & 1.92 & -.01  \\
\end{tabular}
\end{table}

$(1 + z_Q)/(1 + z_G) = (1 + z_0)$

$(1 + z_O)/(1 + z_K) = (1 + z_v)$

G = galaxy, K = Karlsson, Q = quasar

$z_0$ = transformed Q to G

$z_v$ = $z_Q$ seen from Galaxy

$^{a}$ The highest $z_v$ of the first 9 of 10 quasars is $.0165$.

$^{b}$ Figures 1 and 3 were made up from an early version of
NED data. Later two additions were made from SDSS, quasars of $z = 1.807$ and $.702$
having $z_v = +.097$ and $+.003$. Table 1 shows these new entries. The
probabilities were calculated from Table 1. The difference in the
cumulative means of Figure 1 is not significant and is
consistent with considering the $z = 1.807$ as a non associated
interloper or a $z$ value falling briefly between two peaks.

\begin{figure}[ht]
\begin{center}
\includegraphics[scale=0.8, angle=0]{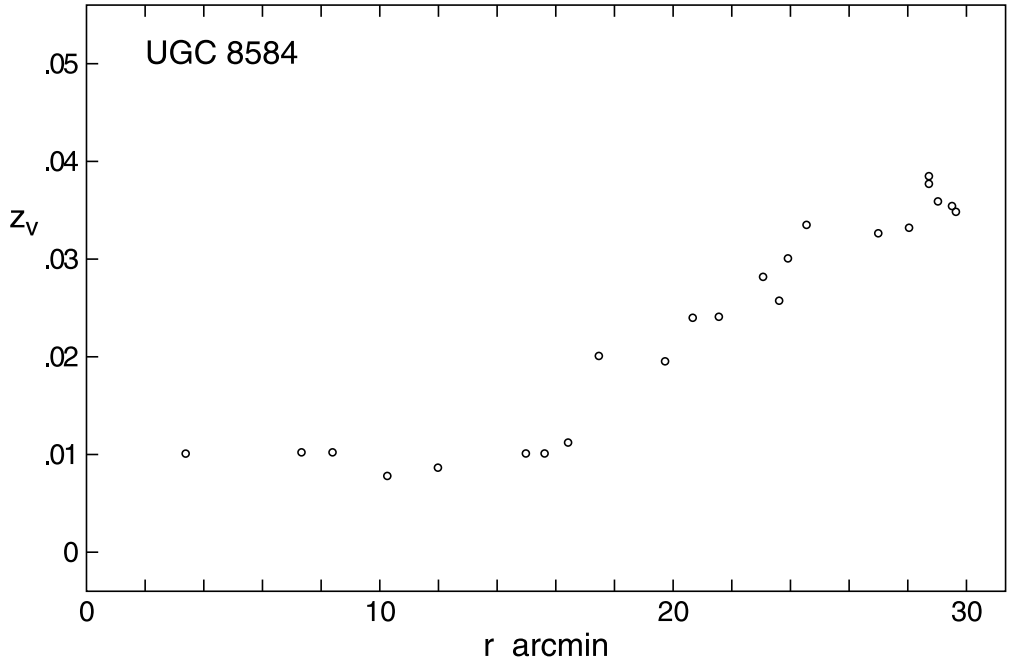}
\caption{$z_v$'s represent cumulative average residuals from Karlsson
periodicity peaks. Of the first 10 quasars out to $r = 16.5'$, 9 are
very close to Karlsson periodicities. See footnotes to Table 1.}
\end{center}
\end{figure}

\begin{figure}[ht]
\begin{center}
\includegraphics[scale=0.8, angle=0]{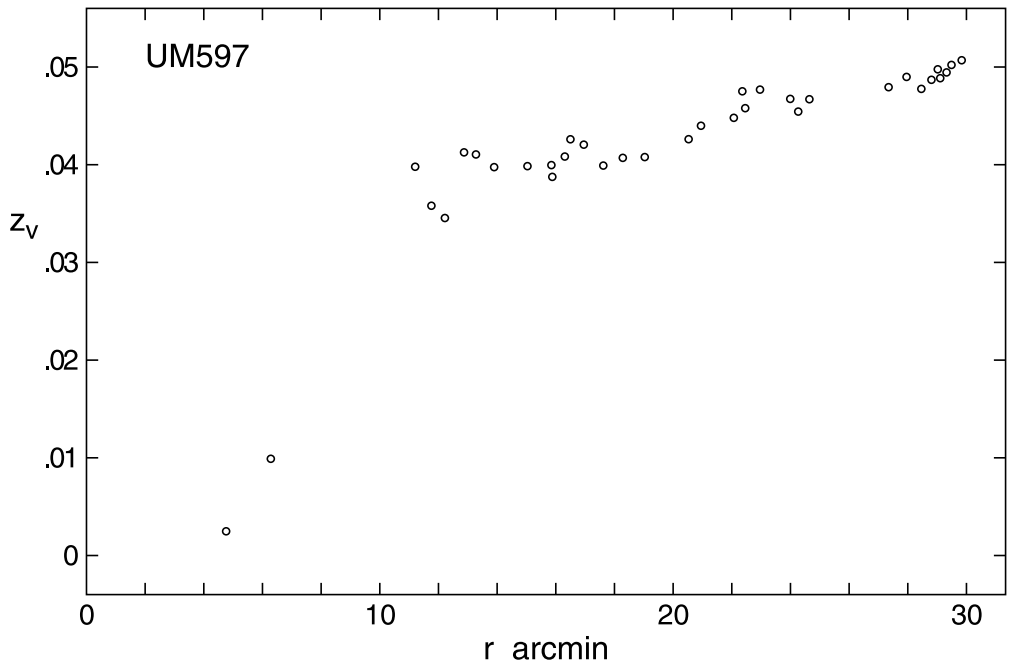}
\caption{As above, $z_v$'s represent cumulative residuals from Karlsson
periodicity peaks. The first two quasars are very close to
this particular parent and to the Karlsson peak at z = 1.96.}
\end{center}
\end{figure}

\clearpage

To make a visual comparison with another field, Figure 2 is shown just
below UGC 8584. In the case of the active galaxy UM 597 the two nearest
quasars are very closely periodic in redshift ($z_0 = 1.97$ and $1.91$)
and within $6.7'$ in distance. But then a large number of quasars set in
with large cumulative $z_v$ averages at greater than about $12'$. In
both plots if the quasars were not near their periodic redshift values
the lines of points would run straight across at a cumulative average
of about $z_v = .0525$. (With larger scatter at small r.) The actual
plots show the effect of quasars matching the periodic values strongly
close to the parent galaxies, but still affecting the plots out past $r = 30'$.

\section{Evidence for Ejection}

Another way of analyzing the periodicity residuals for the quasars
within $30'$ of UGC 8584 is shown in Figure 3. The histogram shows a
symmetrical distribution around zero. There is a strong excess within
$z_v = \pm.03$. If the expected background is taken from the wings of
the histogram this indicates a peak of about 4 sigma significance.

Also important, however, is the fact that the largest concentration of
residuals is at $-.01$ and again at $+.01$. This would be strong evidence
for an ejection velocity of the order of $v = .01c$ with equal numbers
toward and away from the observer. It would be difficult to imagine a
more direct proof of ejection.

\begin{figure}[ht]
\begin{center}
\includegraphics[scale=1.0, angle=0]{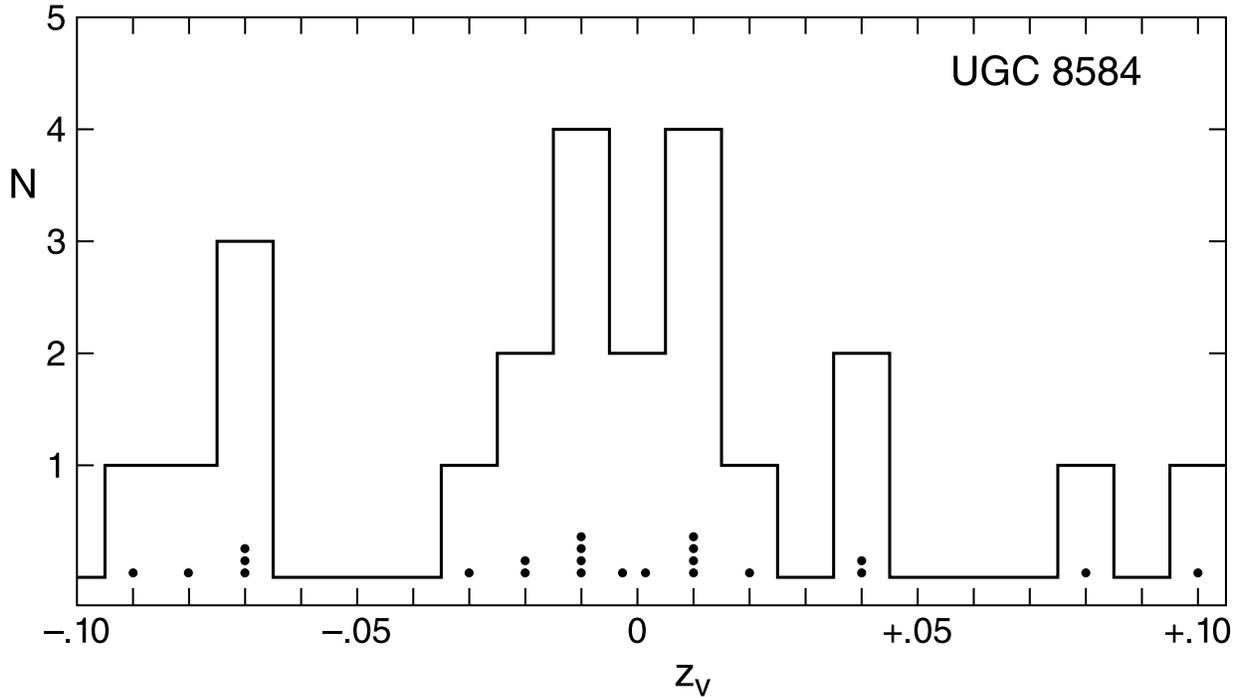}
\caption{Distribution of periodicity residuals, $z_v$.}
\end{center}
\end{figure}

\section{The Disturbed Morphology of UGC 8584}

UGC 8584 is a distorted triple system of about 15 mag. as shown here
in Figure 4. All three objects have $z = .060$. In the high resolution
FIRST measures each of these is shown to be a compact
radio source  with the central one the brightest. It is difficult to
escape the conclusion that the two flanking objects are portions of
the central galaxy torn off or entrained in an ejection event.

\begin{figure}[ht]
\begin{center}
\includegraphics[scale=0.8, angle=0]{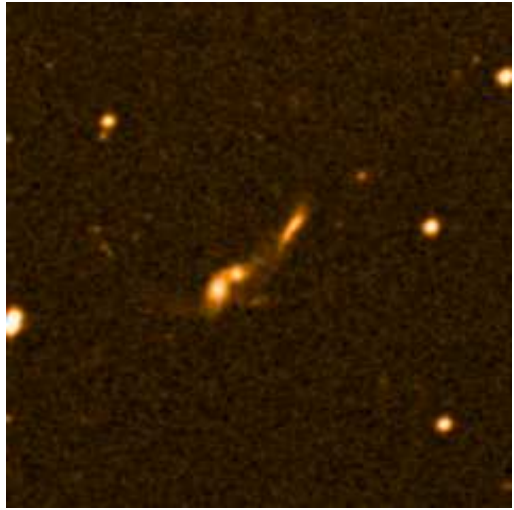}
\caption{Digital Sky Survey blue image of triple system UGC 8584.}
\end{center}
\end{figure}

\begin{figure}[ht]
\begin{center}
\includegraphics[scale=0.5, angle=0]{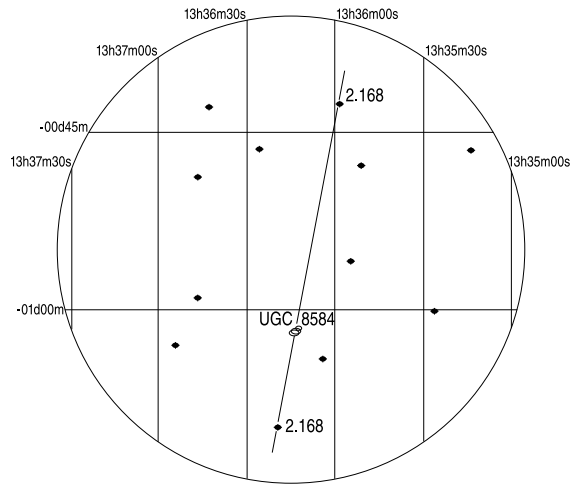}
\caption{Quasars within $20'$ of UGC 8584 showing pair at $z = 2.168$.}
\end{center}
\end{figure}

\newpage

\section{Evidence for Radio Ejection from UGC 8584}

The relation between ejection of radio sources and ejection of quasars
has long been apparent (e.g. Arp (1966), Arp (2003)). In the case of UGC 8584 the
fact that the central galaxy in the association of quasars was bright
and violently disturbed supports the inference of origin from, and
physical association with, UGC 8584. As Figure 6 here shows, UGC 8584 is
also itself a strong radio source, reinforcing the aspect of activity of
the source of the associated quasars.

But something else of great interest appears in Figure 6. UGC 8584 seems
to be in the center of a ring of bright radio sources. There are radio
sources outside this ring but the interesting feature is that radio
sources appear to be cleared out of a circle of radius about $13'$ around
UGC 8584. If there has been violent ejection from the central galaxy it
is reasonable to suppose that most of the low density radio plasma has
been pushed outward leaving a radio source free cavity.

\begin{figure}[ht]
\begin{center}
\includegraphics[scale=0.9, angle=0]{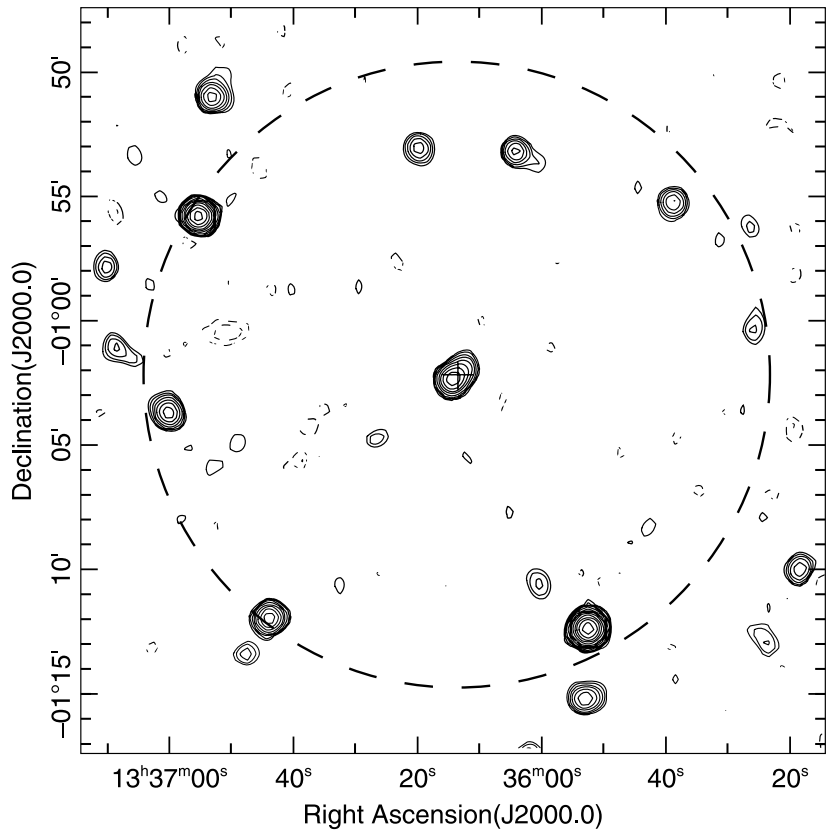}
\caption{NVSS radio map centered on UGC 8584. Dashed circle = $13'$ radius.}
\end{center}
\end{figure}

\newpage

\section{Obscuring Material Ejected?}

It is clear from Figure 1 that inside about $20'$ from UGC 8584 that
the quasars all fall extremely close to Karlsson redshift
peaks. Figure 7 here shows this in a sightly different way that
emphasizes the result that inside about $20'$ there are almost no
quasars that do not obey the Karlsson Peaks. This is strong support
for the physical association of these quasars with UGC 8584. But the
question then rises: where is the background of non-associated
sources?

On average for 2dF quasars we expect 33 QSOs/sq.deg. But Figure 7 shows
that for non associated quasars near UGC 8584 the density is low,
between 3 and 6/sq.deg. By $37'$ the density above $z_v \geq .01$ in
Figure 7 already reaches about 27/sq.deg. It seems  strongly indicated
that a background of quasars near UGC 8584 is being obscured. Although
this should be checked, there seems to be no conspicuous reddening
involved. If so the particle sizes would be larger than dust, perhaps
gravel, boulder or larger sizes.

\begin{figure}[ht]
\begin{center}
\includegraphics[scale=1.3, angle=0]{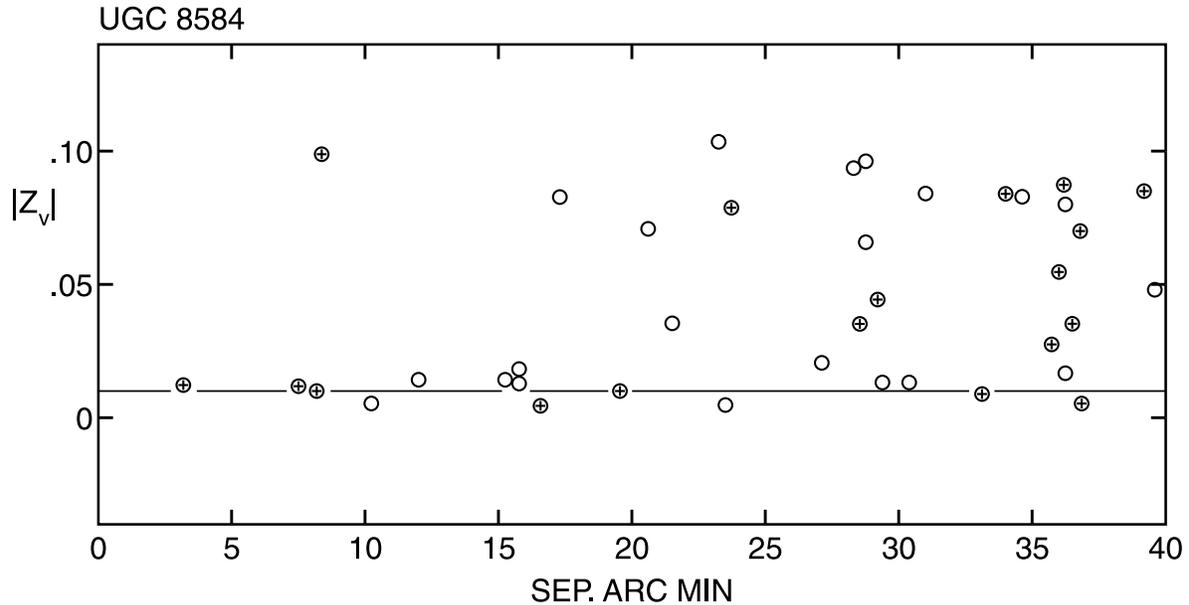}
\caption{Residuals from Karlsson Peaks for redshifts of quasars $(z_v)$
as a function of separation from UGC 8584. Plus and minus residuals
labeled.  Note that the almost exact correspondence with periodicity
peaks is NOT followed by quasars at greater radial distances.}
\end{center}
\end{figure}

Altogether with nine of the nearest ten quasars matching so closely
the periodicity peaks, the disrupted, active nature of the central
galaxy and the ring of radio sources around it, it would seem
difficult to avoid the conclusion that 9 of the nearest 10 quasars are
physically associated with, and probably ejected from, UGC 8584. The
symmetry of the plus and minus residuals in Figure 3 most simply reflects
the remains of the original velocities of ejection in opposite
directions.

That the ejection takes place in a ring or shell is now supported in
several other cases which we briefly discuss.

\newpage

\section{The Ring of Objects Around ESO413-007}

It is helpful that there is another parent galaxy in the current sample
that has a similar, confirming perimeter of radio sources around it.
(This further example was discovered in the same test sampling of a
dozen or so of bright galaxy/quasar families as found in the
computer analyses of this paper series). Figure 8 shows that there is a ring
of objects around the central galaxy ESO413-007 -  objects which turn
out to be both galaxies and quasars of different redshift values.

\begin{figure}[ht]
\begin{center}
\includegraphics[scale=0.6, angle=0]{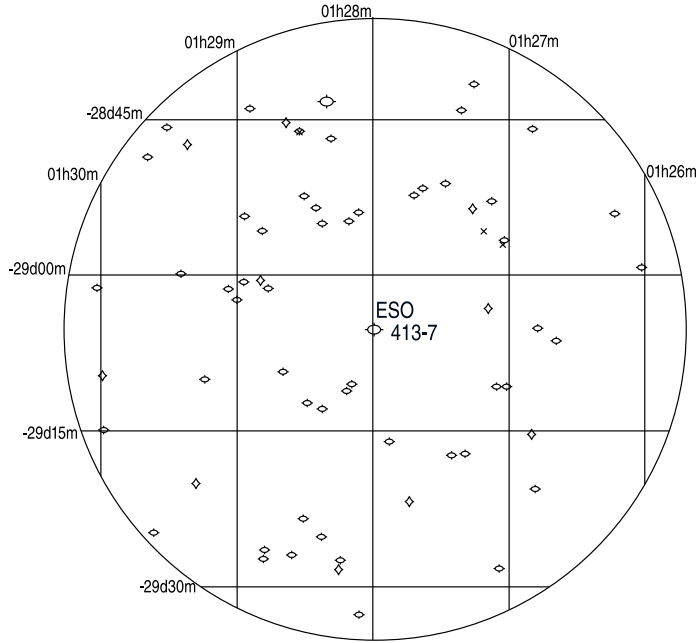}
\caption{Simbad map of extragalactic objects (both galaxies and
quasars) within $30'$ of ESO413-7.}
\end{center}
\end{figure}

\begin{figure}[ht]
\begin{center}
\includegraphics[scale=1.1, angle=0]{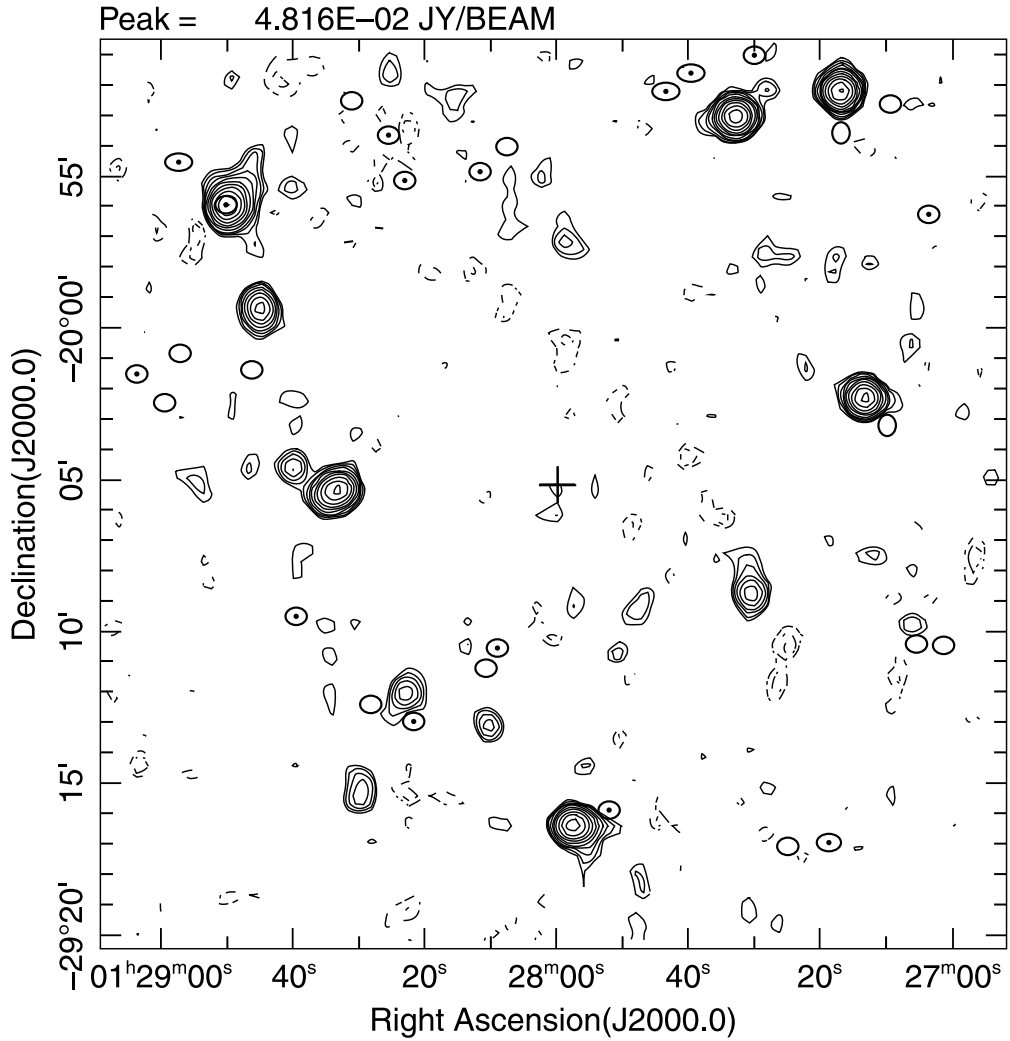}
\caption{NVSS radio map in $30 \times 30'$ square around ESO413-7. Ovals
represent galaxies and ovals with dots in center represent quasars.}
\end{center}
\end{figure}

Figure 8 also shows that the immediate area around the bright ($15.8$ mag.)
infrared galaxy ESO 413-7 is devoid of both galaxies and quasars.
To see the radio sources we refer to
Figure 9. Like UGC 8584 (compare Figures 6 and 9) there is a loose ring of
bright radio sources around the central galaxy and an empty region bordered
by radio and optical sources. In order to demonstrate this ring of
mixed radio and optical sources we have plotted together in Figure 9 the
objects classified as galaxies as ovals and the QSO's as ovals with a
dot in the center. Why do they confirm the same hollow ring pattern
as the radio sources?

What Figures 5 - 9 suggest is that the ejection activity of the central
galaxy has blown a lower density cavity around itself. When the
ejected quasar/plasmoids hit the higher density ring or shell the
lower density radio plasma is slowed or stopped and in any case
stripped from the optical quasar. The optical quasar remains to evolve
into a medium redshift galaxy relatively near the observed
edge. (These higher redshift companions may also eject radio sources
intermittently.)

An interesting side note to Figure 9 is that the nearest quasars (SE of
the central ESO galaxy) do not fall at Karlsson Peaks in its $z = .005$
rest frame. But the nearest NED galaxies, SE, at $18.06$ and $18.35$ mag
are infrared (IrS) sources, only $3.7'$ and $4.6'$ away and just the kinds
of galaxies preferentially found as parents of families of associated
quasars. As Table 2 shows, for galaxies at $z = 0.101$ and $0.132$,
five of these quasars fall exceptionally close to the Karlsson peaks!

\begin{table}[ht]
\caption{Quasars Near ESO 413-007} \label{Table2} \vspace{0.3cm}
\begin{tabular}{lccc}
{\bfseries Object\/} & {\bfseries $z$\/} & {\bfseries $z_0$\/} & {\bfseries $z_v$\/} \\
\hline
compn1 & 0.101 &  ...  &  ...  \\
2QZ    & 1.675 & 1.430 & +.008 \\
2QZ    & 1.186 &  .985 & +.013 \\
2QZ    & 1.168 &  .969 & +.005 \\
$$     &       &       &       \\
compn2 & 0.132 &  ...  &  ...  \\
2QZ    & 2.374 & 1.981 & +.007 \\
2QZ    & 1.186 &  .931 & -.015 \\
\end{tabular}
\end{table}

\clearpage

\section{Small Ring of Optical Galaxies}

A serendipitous example of a ring of optical galaxies is shown in
Figure 10. In examining a group of galaxies NW of NGC 4410 (Arp et
al. 2007) an SDSS picture of an $18.4g$, $z = .089$ galaxy was
encountered. As is in Figs. 6 and 8, there is an almost complete circle
of galaxies, particularly faint ones, at a radius of about
$0.9'$. Because there has been no systematic search for such examples,
and in view of the evidence for rings in the previous two cases just
discussed we would have to conclude there could be many further
examples to be discovered. In this case the the two pairs of
diametrically opposed blue objects have not been checked with
spectra. It would be important if they turned out to be higher
redshift AGNs.

\begin{figure}[ht]
\begin{center}
\includegraphics[scale=0.55, angle=0]{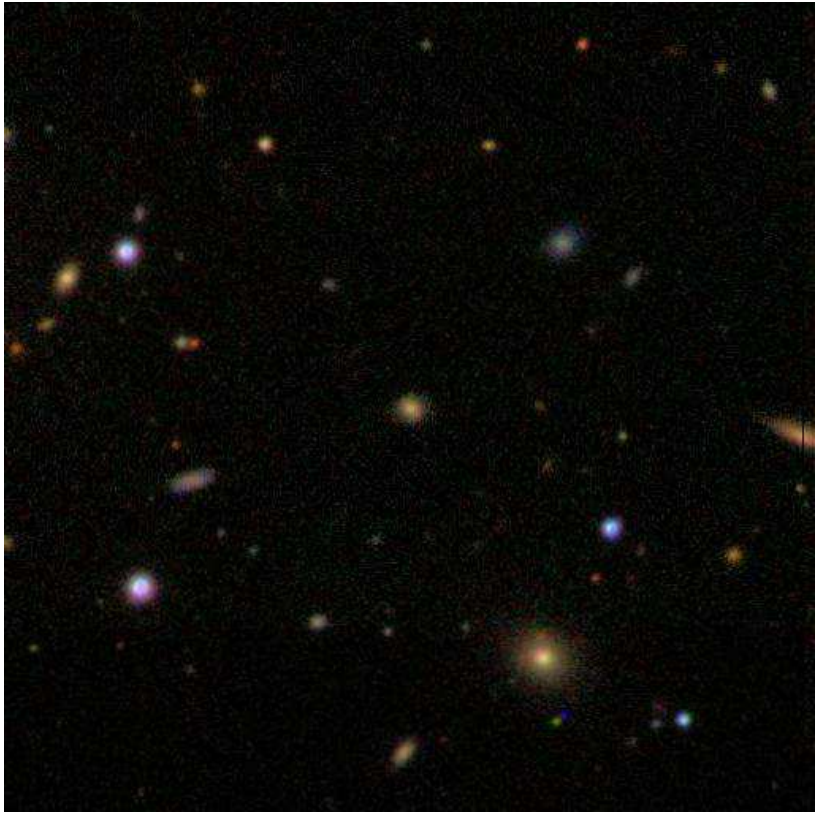}
\caption{An SDSS image of a $3.4'$ field around SDSS 122524.87+092307.1. 
The ring of faint objects has a radius of about $0.9'$ around the $18.4g$
mag. galaxy.}
\end{center}
\end{figure}

\section{Ejection Pairing}

Figure 5 shows the Simbad map of the quasars around UGC 8584. The
most striking feature is the pair of quasars of exactly the same red
shift $z = 2.168$ paired as exactly as can be determined across the
central distorted triplet. Certainly this speaks for an ejection
origin as so many other equal pairs across active galaxies (vide Arp
220.) Also there seems to be a ring of radio sources about $15'$
radius around UGC 8584 as shown in Figure 6.

\newpage

\section{Summary}

In a preliminary sampling, one of the quasar families found in the
periodicity analysis of Paper I shows nine out of ten of the
quasars out to $17'$ to be associated with UGC 8584. When transformed to
the rest frame of the disturbed triple radio galaxies that make up
UGC 8584, they have redshifts that fall unusually close to the
Karlsson peak redshifts. This is like a key fitting into a lock.

Analysis of the residuals from the exact Karlsson peaks shows a
symmetrical balance of plus and minus residuals at about $|z_v| =
.01c.$ This points strongly to their ejection velocity toward and away
from the observer at this point in their evolution to smaller
intrinsic redshifts.

A ring of radio sources around UGC 8584 at about $r = 13'$ suggests a
shell of material has been blown out of the central galaxy during the
ejection of the quasars. Another quasar family, UM 597, supports this
ring configuration of radio sources and suggests that quasar evolution
into galaxies takes place at or near these perimeters. (See e.g.
``Origin of Companion Galaxies'' in Arp (1998a)).

In turn this suggests that the so called gravitational arcs could
represent optical shells that were ejected along with the quasars.
Arguments have been made (Arp (1998b); Arp (2003)) that such optical,
explosion related, arcs have already been observed.\\

\section{Acknowledgments}

This research has made use of the SIMBAD database, which is operated at CDS, Strasbourg, France.

This research has made use of NED, which is operated by the Jet Propulsion Laboratory, California Institute of Technology, under contract with the National Aeronautics and Space Administration.
\\
\\

{\bf References}

Arp, H., 1966, Sci 151, 1214.

Arp, H., 1998a, ApJ 496, 661.

Arp, H., 1998b, ``Seeing Red'' Apeiron, Montreal.

Arp, H., 2003, ``Catalogue of Discordant Redshift Associations.'' Apeiron, Montreal.

Arp, H., Burbidge, G., \& Carosati, D. 2007, ``Quasars and Galaxy
Clusters Paired Across NGC 4410,'' ApJ submitted.

Fulton, C. \& Arp, H., 2006, ``The 2dF Redshift Survey I: Physical
Association and Periodicity in Quasar Families,'' ApJ submitted (Paper I).

Karlsson, K., 1971, A\&A, 13, 333.
    
\end{document}